\documentstyle[aps,epsfig]{revtex}

\begin{document}
\draft
\title{Is the Unitarity of the quark-mixing-CKM-matrix violated in neutron $\beta$-decay?}
\author{H.Abele$^{1}$, M.Astruc Hoffmann$^{1}$,
S.Bae$\ss$ler$^{1}$, D.Dubbers$^{2}$, F. Gl\"uck$^3$,
U.M\"uller$^{1}$, V.Nesvizhevsky$^{2}$, J.Reich$^{1}$,
O.Zimmer$^{2}$} \address{$^{1}$ Physikalisches Institut der
Universit\"at Heidelberg, 69120 Heidelberg, Germany\\ $^{2}$
Institut Laue-Langevin, B.P. 156, F-38042 Grenoble Cedex 9,
France\\ $^{3}$ KFKI RMKI, 1525 Budapest, POB 49, Hungary}
\date{\today} \maketitle \begin{abstract}
\date{\today}
\maketitle
\begin{abstract}
We report on a new measurement of neutron $\beta$-decay asymmetry.
From the result \linebreak $A_0$ = -0.1189(7), we derive the ratio
of the axial vector to the vector coupling constant $\lambda$ =
${\it g_A/g_V}$ = -1.2739(19). When included in the world average
for the neutron lifetime $\tau$ = 885.7(7)s, this gives the first
element of the Cabibbo-Kobayashi-Maskawa (CKM) matrix $V_{ud} $.
With this value and the Particle Data Group values for $V_{us}$
and $V_{ub}$, we find a deviation from the unitarity condition for
the first row of the CKM matrix of $\Delta$ = 0.0083(28), which is
3.0 times the stated error. \end{abstract}
\end{abstract}
 \narrowtext \twocolumn

As is well known, the quark eigenstates of the weak interaction do
not correspond to the quark mass eigenstates. The weak eigenstates
are related to the mass eigenstates in terms of a 3 x 3 unitary
matrix $V$, the so called Cabibbo-Kobayashi-Maskawa (CKM) matrix.
By convention, the u, c and t quarks are unmixed and all mixing is
expressed via the CKM matrix ${\it V}$ operating on d, s and b
quarks. The values of individual matrix elements are determined
from weak decays of the relevant quarks. Unitarity requires that
the sum of the squares of the matrix elements for each row and
column be unity. So far precision tests of unitarity have only
been possible for the first row of ${\it V}$, namely
\begin{equation} |V_{ud}|^2 + |V_{us}|^2 + |V_{ub}|^2 = 1-\Delta.
\end{equation} In the Standard Model, the CKM matrix is unitary
with $\Delta$ = 0. Usually, $|V_{ud}|$ is derived from
superallowed nuclear $\beta$-decay experiments to $|V_{ud}|$ =
0.9740(5). This value includes nuclear structure effect
corrections. Combined with kaon-, hyperon- and B-decays, this
leads to $\Delta$ =0.0032(14), signaling a deviation from the
unitarity condition by 2.3 $\sigma$ standard deviation
\cite{Hardy}. However, some of the nuclear corrections are
difficult to calculate, and therefore the Particle Data Group
\cite{Groom} doubles the error in $|V_{ud}|$.

A violation of unitarity in the first row of the CKM matrix is a
challenge to the three generation Standard Model. The data
available so far do not preclude there being more than three
generations; CKM matrix entries deduced from unitarity might be
altered when the CKM matrix is expanded to accommodate more
generations \cite{Groom,Marciano1}. A deviation $\Delta$ has been
related to concepts beyond the Standard Model, such as couplings
to exotic fermions \cite{Langacker1,Maalampi}, to the existence of
an additional Z boson \cite{Langacker2,Marciano2} or to the
existence of right-handed currents in the weak interaction
\cite{Deutsch}. Non-unitarity of the CKM matrix in models with an
extended quark sector give rise to an induced neutron electric
dipol moment that can be within reach of the next generation of
experiments \cite{Liao}.

In this article, we derive $|V_{ud}|$, not from nuclear
$\beta$-decay, but from neutron decay data. In this way, the
unitarity check of (1) is based solely on particle data, i.e.
neutron $\beta$-decay, K-decays, and B-decays, where theoretical
uncertainties are significantly smaller. So much progress has been
made using highly polarized cold neutron beams with an improved
detector setup that we are now capable of competing with nuclear
$\beta$-decays in extracting a value for $V_{ud}$, whilst avoiding
the problems linked to nuclear structure.

In the Standard Model, the Lagrangian of neutron decay is
restricted to:
\begin{eqnarray}
{\cal L}_{int}&=&\frac{G_F}{\sqrt{2}}{V_{ud}}\cdot
\{\bar{p}[\gamma^\mu(1+\lambda\gamma_{5})\nonumber\\
&&{}+\frac{\mu_p-\mu_n}{2m_p}
\sigma_{\mu\nu}q^\nu]n\cdot\bar{e}\gamma^\mu(1-\gamma_{5})\nu\}.
\end{eqnarray}
$G_F$ is the Fermi decay constant; $n$, $p$, $e$ and $\nu$ are
spinors describing neutron, proton, electron and neutrino;
$\lambda$ is the ratio of the axial vector to the vector coupling
constant $g_A/g_V$; and $q$ is the momentum transfer between
hadrons and leptons. The term ($\mu_p-\mu_n$)/2$m_p$ is the weak
magnetism contribution, which is linked to $\mu_p$ and $\mu_n$,
the anomalous magnetic moments of proton and neutron. $m_p$ is the
nucleon mass. In Eq. (2), we omitted the second-class induced
scalar form factor $f_3$ and the induced pseudotensor form factor
$g_2$, because second class currents are excluded in the Standard
Model. The first-class induced pseudoscalar form factor $g_3$ is
negligible in neutron decay and constrained by the
Goldberger-Treiman relation.

Since $G_F$ is known from muon decay, in the Standard Model only
two additional parameters are needed to describe free neutron
decay, namely $\lambda$ and $V_{ud}$. In principle, the ratio
$\lambda$ can be determined from QCD lattice gauge theory
calculations, but the results of the best calculations vary by up
to 30\%. Today therefore all weak semileptonic particle
cross-sections used in cosmology, astrophysics and particle
physics have to be calculated from neutron decay data.

A neutron decays into a proton, an electron and an electron
antineutrino. Observables are the neutron lifetime $\tau$ and
spins $\sigma_e$, $\sigma_\nu$, $\sigma_p$, and momenta $p_e$,
$p_\nu$, $p_p$ of the electron, antineutrino and proton
respectively. The electron spin, the proton spin and the
antineutrino are not usually observed. The lifetime is given by
\begin{equation}
\tau^{-1}=C |V_{ud}|^2(1+3\lambda^2) f^R(1+\Delta_R),
\end{equation}
where $C=G_F^2 m_e^5/(2 \pi^3)=1.1613\cdot10^{-4} s^{-1}$ in
$\hbar=c=1$ units, $f^R$ =1.71482(15) is the phase space factor
\cite{Wilkinson1} (including the model independent radiative
correction) adjusted for the current value of the neutron-proton
transition energy.
%Natural constants are subsumed in $k$ =
%1.775$\cdot$10$^{-120}$J$^2$m$^6$s$^{-1}$.
$\Delta_R$ = 0.0240(8) \cite{Hardy,Towner1} is the model dependent
radiative correction to the neutron decay rate, of which 0.0212 is
straightforward electroweak-asymptotic QCD contribution, whereas
the remaining 0.0028 depends on the strong interaction models. The
neutron $\beta$-decay rate and its relevant uncertainties at the
10$^{-4}$ level were reviewed recently \cite{Garcia,Bunatian}.

The probability that an electron is emitted with angle $\vartheta$
with respect to the neutron spin polarization \linebreak $P$ =
$<\sigma_z>$ is \cite{Jackson} \begin{equation} W(\vartheta) = 1
+\frac{v}{c}PA\cos(\vartheta), \end{equation} where $v/c$ is the
electron velocity expressed in fractions of the speed of light.
${\it A}$ is the $\beta$-asymmetry coefficient which depends on
$\lambda$. On account of order 1\% corrections for weak magnetism,
$g_V-g_A$ interference,
and nucleon recoil, ${\it A}$ has the form $A$ = $A_0$(1+$A_{{\mu}m}$($A_1W_0+A_2W+A_3/W$)) %\begin{equation} %A = A_0(1+a_{{\mu}m}(A_1W_0+A_2W+A_3/W)) %\end{equation}
with electron total energy $W = E_e /m_ec^2+1$ (endpoint $W_0$).
$A_0$ is a function of $\lambda$
\begin{equation}  A_0=-2\frac{\lambda(\lambda+1)}{1+3\lambda^2},
\end{equation} where we have assumed that $\lambda$ is real. The coefficients $A_{{\mu}m}$, $A_1$, $A_2$, $A_3$ are from \cite{Wilkinson1} taking a different $\lambda$ convention into consideration. In addition, a further small radiative
correction \cite{Gluck} of order 0.1\% must be applied. Other
correlation coefficients (not measured in our experiment) are the
antineutrino-electron correlation $a$, the antineutrino-asymmetry
correlation $B$, and the time-reversal-violating triple
correlation coefficient $D$. They also depend on $\lambda$. Hence,
various observables are accessible to experiment, so that the
problem in extracting $\lambda$ and $|V_{ud}|$ is overdetermined
and, together with other data from particle and nuclear physics,
many tests of the Standard Model become possible. Of course, a
pertinent determination of the radiative corrections of Eq. (3)
remains an important task.

In the following section we report on our new measurement of the
neutron $\beta$-asymmetry coefficient ${\it A}$ with the
instrument PERKEO II, and on the consequences for $|V_{ud}|$. Our
first measurement \cite{Abele} with this new spectrometer gave a
value for $A_0$ which differed significantly from the combined
previous data. In the light of this we decided to remeasure
coefficient $A_0$ with an improved setup. The new result confirms
our earlier finding, with a reduced error but same result. PERKEO
II was installed at the PF1 cold neutron beam position at the High
Flux Reactor at the Institut Laue-Langevin, Grenoble. Cold
neutrons are obtained from a 25 K deuterium cold moderator near
the core of the 57 MW uranium reactor. The neutrons are guided via
a 60 m long neutron guide of cross-section 60 x 120 mm$^2$ to the
experiment and are polarized by a supermirror polarizer of 30 x 45
mm$^2$ cross section. The de Broglie wavelength spectrum of the
cold neutron beam ranges from about 0.2 nm to 1.3 nm. Above a
wavelength of $\lambda$ $>$ 1.3 nm, a very high degree of
polarization is difficult to achieve. A long wavelength cut-off
filter just in front of the supermirror polarizer removes these
undesired neutrons from the beam \cite{Hogoj}. The degree of
neutron polarization was measured to be $P$ = 98.9(3)\% over the
full cross-section of the beam. The polarization efficiency was
monitored and it remained constant during the whole experiment.
The neutron polarization is reversed periodically with a current
sheet spin flipper, with measured spin flip efficiency of $f$ =
99.7(1)\%.

The main component of the PERKEO II spectrometer is a
superconducting 1 T magnet in a split pair configuration, with a
coil diameter of about one meter. Neutrons pass through the
spectrometer, whereas decay electrons are guided by the magnetic
field to either one of two scintillation detectors with
photomultiplier readout. The detector's solid angle of acceptance
is truly 2x2$\pi$ above a threshold of 60 keV. Electron
backscattering effects, serious sources of systematic error in
$\beta$-spectroscopy, are effectively suppressed. Technical
details about the instrument can be found in \cite{Reich2}.

The measured electron spectra $N^\uparrow_i(E_e)$ and
$N^\downarrow_i(E_e)$ in the two detectors (i=1,2) for neutron
spin up and down, respectively, define the experimental asymmetry
as a function of electron kinetic energy $E_e$ \begin{equation}
A_{i_{exp}}(E_e)=\frac{N^\uparrow_i(E_e) -
N^\downarrow_i(E_e)}{N^\uparrow_i(E_e) + N^\downarrow_i(E_e)}.
\end{equation}
By using (4) and with $<\cos(\vartheta)>$ = 1/2, $A_i{_{exp}}(E)$
is directly related to the asymmetry parameter
\begin{equation}
A_{exp}(E_e)=A_{1_{exp}}(E_e)-A_{2_{exp}}(E_e)=\frac{v}{c}APf.
\end{equation}

The main experimental errors are due to $\em{neutron\ spin\
polarization, background\ subtraction\ {\em and}}$ $detector\
response$. To analyze the $neutron$\ $spin$\ $polarization$, a
special setup of three additional spin flippers and two
supermirror polarizers was used \cite{Serebrov}. This gave an
uncertainty of 0.3\% in the measured beam polarization. As a very
precise cross-check, in a separate setup, neutron polarization was
measured again with three completely different methods: firstly,
with our supermirror setup, secondly, with a new method using an
almost opaque $^3$He spin filter \cite{Heil,Zimmer1}, and thirdly,
with a polarized proton filter\cite{Zimmer2}. The results of all
three measurements agree to within 0.15\%.

A correction of 0.5\% on ${\it A}$ is due to $background$
$subtraction$. One main feature of the PERKEO II spectrometer is
its high $\beta$-decay count rate of about 270 s$^{-1}$ due to a
large decay volume (80 x 80 x 270 mm$^3$) and a 4$\pi$ detector.
As a consequence, the signal- to-background rate in the range of
interest (Fig. 1) was 7:1. Most of the background is environmental
and was measured separately and subtracted from the data. Extreme
care is required to suppress any beam-related background, as
discussed in \cite{Abele}. The beam divergence was limited to 12
mrad by appropriate neutron baffles upstream, made from enriched
$^6$LiF plates. The beamstop was positioned 6 m downstream of the
decay volume. The $\beta$-detectors were installed far off the
beam at a transverse distance of 960 mm, and had no direct view to
the polarizer, the baffles or the beam stop. Any beam-related
gamma quantum had to undergo multiple directional changes before
reaching the detector. Compared with our previous measurement with
the same apparatus \cite{Abele}, this beam-related background was
reduced by a factor of three to 0.3 $s^{-1}$ or to 1:200 of the
signal rate. Thus, in the fit interval the size of the background
correction is 0.5\%. This beam-related background was determined
using an extrapolation procedure described in \cite{Abele} and
\cite{Reich2}. The relative uncertainty of 50\% is a conservative
estimate of this background extrapolation method (see Table 1).

The $detector\ response\ function$ was determined with six
conversion line sources on 10$\mu$g/cm$^2$ carbon backings, which
were remotely inserted into the spectrometer. The K, L, M and N
conversion electrons and the corresponding Auger electrons are
taken into account. Detector response is linear in energy within
1\% leading to an uncertainty of 0.2\% in ${\it A}$ \cite{Reich2}.

\begin{figure}[h]
\noindent \centering \epsfig{figure=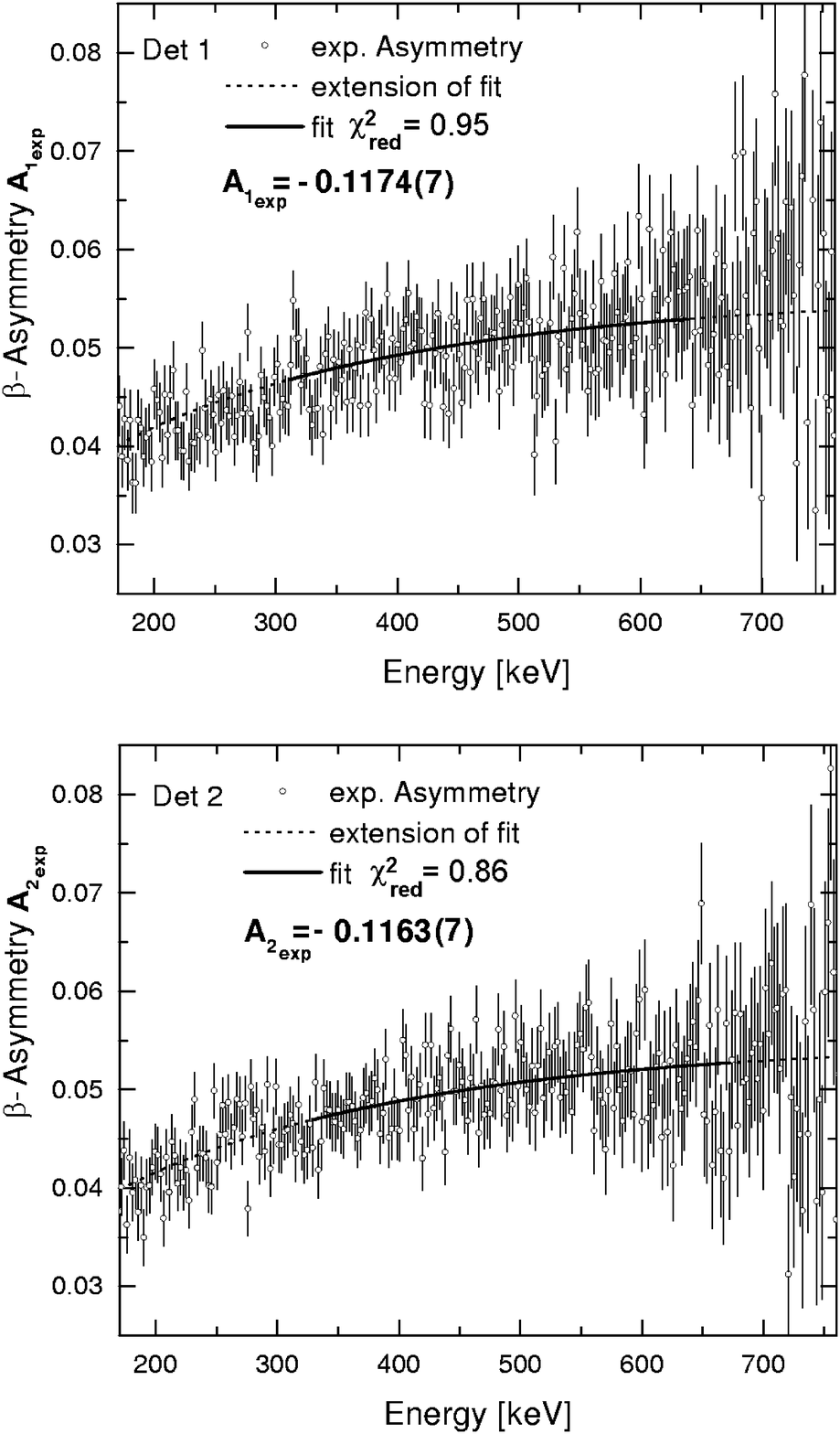, width=8cm}\caption{
Fit to the experimental asymmetry $A_{exp}$ for detector 1 and
detector 2. The solid line shows the fit interval, whereas the
dotted line shows an extrapolation to higher and lower energies.}
\label{fit}
\end{figure}

The experimental function $A_{i_{exp}}(E_e)$ and a fit with one
free parameter $A_{i_{exp}}$ (the absolute scale of $A_0$) is
shown in Fig. \ref{fit}. The $\chi^{2}$ is 142.5 for detector one
and $\chi^{2}$ is 129 for detector two for 150 degree of freedom.
The fit interval was chosen such that the signal-to-background
ratio was at maximum. The parameter $A_{i_{exp}}$ is directly
related to the asymmetry parameter via $A_{i_{exp}}$ =
$A_0$$\cdot$$P$$\cdot$$f$. From the experimental asymmetries we
get $A_{1_{exp}}$ = -0.1174(7) and $A_{2_{exp}}$ = -0.1163(7) for
detector 1 and detector 2 respectively. All corrections and
uncertainties entering the determination of A are listed in Tab.
1. The corrections marked with an asterisk are already included in
the fit. After correcting for the other small experimental
systematic effects listed in Tab. 1, we obtain $A_0$ = -0.1189(8).
This value is identical to our earlier result \cite{Abele} of
$A_0$ = -0.1189(12), but with a smaller error. The combined result
is
\begin{equation} A_0 = -0.1189(7) \mbox{ and } \lambda =
-1.2739(19). \end{equation} With this value, and the world average
value for $\tau$ = 885.7(7) s from \cite{Groom}, we find from (3)
that
\begin{equation} |V_{ud}|=0.9713(13). \end{equation}
With \cite{Groom} $|V_{us}|$ = 0.2196(23) and the negligibly small
$|V_{ub}|$ = 0.0036(9), we obtain \begin{equation} |V_{ud}|^2 +
|V_{us}|^2 + |V_{ub}|^2 = 1 - \Delta = 0.9917(28). \end{equation}
This value differs from the Standard Model prediction by deviation
$\Delta$ = 0.0083(28), or 3.0 times the stated error.

Earlier experiments \cite{Yerozolimsky,Schreckenbach,Bopp} gave
significantly lower values for $|\lambda|$. However, in all these
earlier experiments large corrections had to be made for neutron
polarization, electron-magnetic mirror effects or background,
which were all in the 15\% to 30\% range. In our experiment on the
other hand, the total correction to the raw data is 2.0\%, i.e. 10
times less than in earlier experiments. We therefore believe that
our new experiment is more reliable than previous experiments.

Averaging over our new result and previous neutron $\beta$-decay
results the Particle Data Group \cite{Groom_1} arrives at a new
world average for $|V_{ud}|$ from neutron $\beta$-decay which
leads to only a 2.2 $\sigma$ deviation from unitarity.

The Particle Data Group obtains from superallowed
0$^{+}\rightarrow 0^{+}$ transitions a value $|V_{ud}|$ =
0.9740(10). This value is compatible with our value (9) at the 90
\% C.L..

An independent test of CKM unitarity comes from W physics at LEP
\cite{Sbarra} where W decay hadronic branching ratios can be used.
%expressed in terms of %\begin{equation}
% \frac{Br(W\rightarrow q\bar{q})}{1-Br(W\rightarrow
% q\bar{q})}=(1+\frac{\alpha}{\pi}\sum{|V_{ij}|^2}).
%\end{equation}
Since decay into the top quark channel is forbidden by energy
conservation one would expect $\sum{|V_{ij}|^2}$ to be 2 with a
three generation unitary CKM matrix. The experimental result is
2.032(32), consistent with (10) but with considerably lower
accuracy.

In the frame of the present article we do not want to speculate on
the origin of the deviation $\Delta$. Nevertheless, we want to
point out that it is unlikely that this deviation is due to
induced form factors (as discussed above) or erroneous radiative
corrections or to the other CKM elements $V_{us}$ and $V_{ub}$. A
non-zero second-class term $g_2$ (in contradiction to the Standart
Model) is unlikely because the present experimental limit on $g_2$
$<$ 0.2 \cite{Wilk} would lead to a change in the neutron
decay-asymmetry $A_0$ by less than 0.15\%. If the deviation
$\Delta$ was due to the radiative correction $\Delta_R$, than the
error on $\Delta_R$ must be more than 9 times larger than the
quoted error. Also it is unlikely that a nonzero $\Delta$ is due
to an error in the determination of the high energy results on
$V_{us}$ because the error in $V_{us}$ must be enlarged by factors
of 8 to explain our value of $\Delta$ (the value of $V_{ub}$ is
completely negligible in this context).

In summary, $|V_{ud}|$, the first element of the CKM matrix, has
been derived from neutron decay experiments in such a way that a
unitarity test of the CKM matrix can be performed based solely on
particle physics data. With this value, we find a 3 $\sigma$
standard deviation from unitarity, which conflicts the prediction
of the Standard Model of particle physics.

This work was funded by the German Federal Ministry for Research
and Education under contract number 06HD953.

\newpage
\newpage
\begin{figure}[h]
\noindent \centering \epsfig{figure=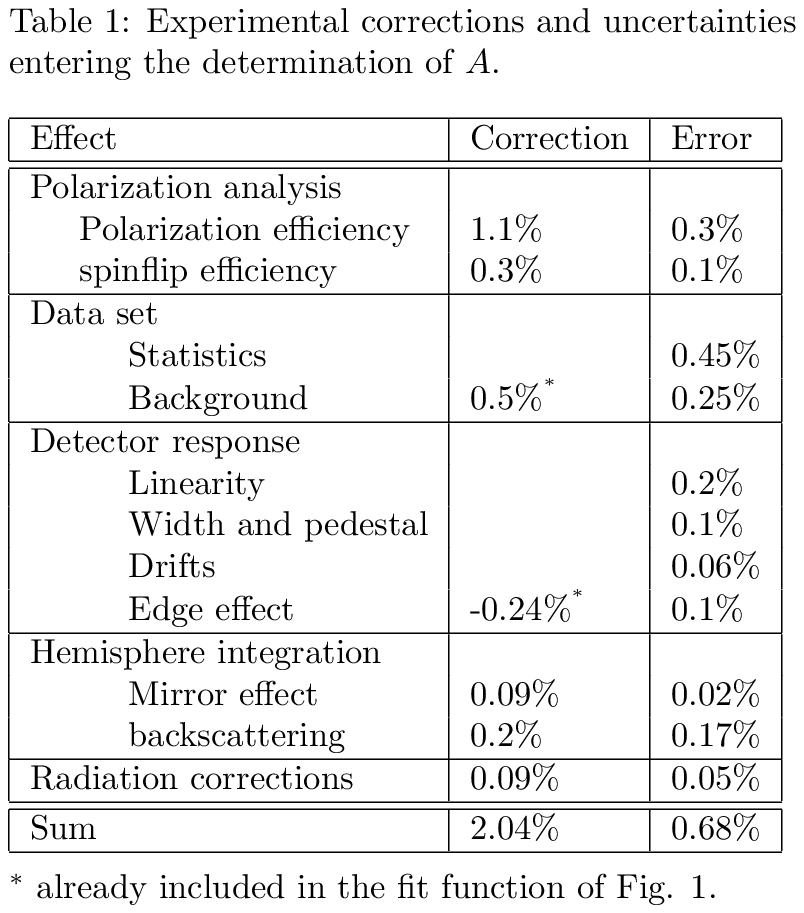, width=8cm}
\end{figure}
\end{document}